# Ion beam lithography for Fresnel zone plates in X-ray microscopy

Kahraman Keskinbora,[1] Corinne Grévent,[1,*] Michael Bechtel,[1] Markus Weigand,[1] Eberhard Goering,[1] Achim Nadzeyka,[2] Lloyd Peto,[2] Stefan Rehbein,[3] Gerd Schneider,[3] Rolf Follath,[3] Joan Vila-Comamala,[4] Hanfei Yan[5] and Gisela Schütz[1]

[1]*Max Planck Institute for Intelligent Systems, Heisenbergstr. 3, 70569 Stuttgart, Germany*
[2]*Raith GmbH, 44263 Dortmund, Germany*
[3]*HZB Bessy II, Albert-Einstein-Str. 15, 15489 Berlin, Germany*
[4]*Advanced Photon Source, Argonne National Laboratory, 9700 South Cass Ave., Argonne, IL 60439, USA*
[5]*National Synchrotron Light Source II, Brookhaven National Laboratory, Upton, New York 11973, USA*
[*]*grevent@is.mpg.de*

**Abstract:** Fresnel Zone Plates (FZP) are to date very successful focusing optics for X-rays. Established methods of fabrication are rather complex and based on electron beam lithography (EBL). Here, we show that ion beam lithography (IBL) may advantageously simplify their preparation. A FZP operable from the extreme UV to the limit of the hard X-ray was prepared and tested from 450 eV to 1500 eV. The trapezoidal profile of the FZP favorably activates its $2^{nd}$ order focus. The FZP with an outermost zone width of 100 nm allows the visualization of features down to 61, 31 and 21 nm in the $1^{st}$, $2^{nd}$ and $3^{rd}$ order focus respectively. Measured efficiencies in the $1^{st}$ and $2^{nd}$ order of diffraction reach the theoretical predictions.

## 1. Introduction

X-ray microscopy successfully combines micro- and nano-scale resolutions with high penetration depths and high chemical sensitivity as well as magnetic imaging capabilities with polarized X-rays. Over the years it has established as one of the preferred methods to characterize the microstructure, the composition or the magnetic structure of a wide variety of entire samples in their natural environment [1-4]. Developments in the last decades have been characterized by the introduction of new X-ray sources of high brilliance [5-13] and by a substantial increase of the achievable spatial resolutions [14-17]. These improvements suggest and support an even expanded utilization of X-ray microscopy in the future, both at large facilities [6] and in laboratories [5, 11]. Nevertheless, the fabrication of corresponding high resolution focusing optics is still complex and challenging.

High spatial resolution imaging to date is best achieved with a diffraction based focusing optics called Fresnel Zone Plate (FZP). FZPs are constituted of a set of concentric rings or zones of decreasing width towards the outside [1]. With more than one hundred zones they behave like simple lenses [18] but present several foci corresponding to the various diffraction order $m$ of the FZP. The FZP's diffraction efficiency decreases with the square of the diffraction order $m$ [19] and depends on various parameters such as the thickness, the constituting material, the profile of the zones and the employed radiation energy. The FZP's spatial resolution is related to the width of the outermost zone (Δr) and increases linearly with the diffraction order $m$ according to equation (1).

$$\delta_{Ray}^m = \frac{1.22\Delta r}{m} \tag{1}$$

where equation (1) valid for a fully incoherent imaging process.
Up to now the favored strategy towards high resolution has consisted in fabricating FZPs with small Δr. In practice this is a highly challenging nano-structuring issue where electron beam lithography (EBL) based methods have established as standard [1]. In the extreme UV and soft X-ray range they allow for FZPs resolving 20 to 15 nm features and even smaller [20-22]. However, thought very successful EBL based fabrications are rather complex and typically involves at least 4 different fabrication steps [1, 20-22].

Here, we show that ion beam lithography (IBL) can be used as a simple single-step method to efficiently prepare Fresnel Zone Plates for a wide range of wavelengths ranging from the extreme UV to the limit of hard X-rays. A gold-FZP with a diameter of 100 μm, a thickness of 500 nm and a Δr of 100 nm has been prepared and characterized. The resolution in the first diffraction order reaches the expected Rayleigh resolution and 61 nm features could clearly be observed. As the method inherently leads to the fabrication of FZPs with trapezoidal profiles, the 2$^{nd}$ diffraction order is efficiently activated and following equation (1) , resolutions twice as high as in the 1$^{st}$ order of diffraction were achieved. Structures of 31 nm in size were resolved with good quality and reproducibility. Moreover the achieved fabrication accuracy is high enough to allow for good imaging properties up to the 3$^{rd}$ diffraction order. Structures down to nearly 21 nm in size (half pitch cutoff) could also be resolved.

2. **Ion beam lithography as manufacturing technique for Fresnel Zone Plates**

Over the past decades, alternatives to the well known EBL based technologies for the fabrication of FZPs were conceived and tested with the aim of either simplifying the preparation, enhancing the resolutions or producing FZPs for shorter wavelength. Multilayer FZPs [23-24] were for instance reported [25] to achieve higher resolutions at shorter wavelengths and higher throughput. As for the simplification of the preparation ion beam lithography [26-27] (IBL) based technologies were also envisaged. They would advantageously permit the fabrication of FZPs in one single-step. This would be interesting in many issues, as a simplification and hence a reduction of the number of required skills for the overall FZP fabrication and a reduction of the typical accumulation of errors characteristic of multi-step processes should in any case allow a faster easier and precise fabrication. As another advantage, the method requires no additional metallic layers, which are usually employed as plating base in EBL for the further electroplating process and which reduces the overall efficiency of the FZP especially at lower energy. Despite those advantages, IBL had received only little attention to date. First attempt of FZP preparation proved to be difficult and till very recently the imaging properties of the prepared FZPs had not been tested [28-30]. Nevertheless recently, we could show that FZPs with good imaging properties can indeed be prepared with IBL and features as small as 120 nm could be resolved in the soft X-rays range at 1200 eV [31]. Other IBL prepared FZPs with resolutions of about 172 nm and their utilization as objective lens in a full field laboratory microscope in the extreme UV range at 13 nm wavelength (95 eV) were also reported a few months later [32].

3. **Fabrication and Design**

Two identical FZPs were fabricated and tested. The FZPs were prepared in a polycrystalline 500 nm thick gold layer deposited via ion beam sputtering on a 500 nm thick Silicon Nitride Membrane (Silson Ltd., England). For the fabrication a new IBL system was used (*ionLINE*®, Raith Germany). This IBL tool is a recent advancement over other conventional focused ion beam (FIB) instruments. It uses an advanced 3D variant patterning software, a laser interferometer stage, up to date ion sources as well as new ion focusing optics providing enhanced emission stabilities and spot shapes. With optimized drift corrections and the enhanced beam stability it allows the milling of very small structures even at large ion beam incident angles which enable the fabrication of FZPs with large diameters. As IBL parameters, we used an ion beam energy of 40 keV, an ion beam current of 50 pA, a radiation dose of 0.35 nC/µm² and a process time of 15 hours. The precise description of the IBL fabrication process can be found elsewhere [31]. The placement of the absorbing gold zones follows the standard FZP formula [1] for a wavelength of 1.03 nm (energy 1200 eV), an outermost zone width of 100 nm and a total diameter of 100 µm. The achieved aspect ratio is thus of 5.

Compared with standard FZPs with a rectangular profile [1], the present FZPs nevertheless present some particularities. Due to typical redeposition processes in IBL [33] the zones of the FZP present a trapezoidal profile (see Fig. 1 lower part) with slanted walls characterized by angles varying from $\alpha = 4°$ to $6°$ from the inner part to the outer part of the FZP. The inclination of the walls can be adjusted by varying the ion beam current. The use of lower beam currents potentially allows a reduction of $\alpha$ down to about $2°$ [28]. The FZP was fabricated without center stop. Additionally, it is a positive FZP, which means that the first and the last zones are open [34] while in our case the first zone was left full. [Fig. 1]

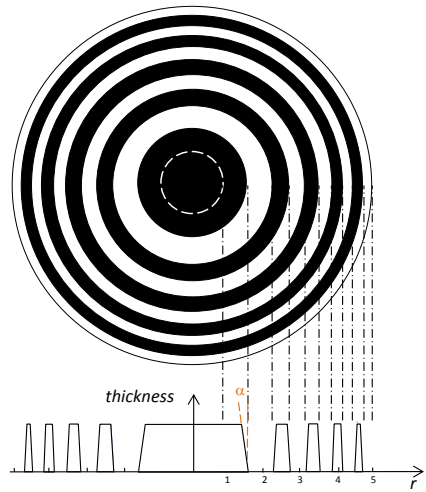

Fig. 1. Representation of the fabricated gold FZP (up and cross-sectional side view): Only a few zones are displayed for sake of simplicity. Positive FZP for which the first zone is filled with gold and the last zone is empty; the zone material is polycrystalline gold; the zone height is 500 nm; the diameter is 100 µm; the outermost zone period Λ is 200 nm (corresponding to an outermost zone width Δr of 100 nm); the number of zones N is 251; α is the angle characteristic of the inclination of the walls as a result of ion beam lithography.

The ability of a FZP to focus a large range of wavelengths is essentially determined by its thickness and to a lesser extent by chromaticity issues. High FZP thicknesses are crucial because they are necessary to the efficient focusing of high energy radiation. Nevertheless, they are difficult to realize with typical lithography processes, for which the achievement of high aspect ratio is difficult. Here the achieved gold thickness of 500 nm allows good theoretical efficiencies between 7.3% and 21.9% for energies between 30 eV and 8000 eV (Table 1). These efficiencies, were calculated within the framework of the thin grating approximation following the approach proposed by Kirz [19] for trapezoidal FZPs by considering a placement of the zones following the standard design [1] and an angle α [Fig. 1] characterizing the trapezoid of 6°.

Table 1 Calculated $1^{st}$ order diffraction efficiencies for a trapezoidal FZP made out of a 500 nm thick gold layer, $\Delta r = 100$ nm and $\alpha=6°$ within the frame work of the thin grating approximation [19] which is valid as long as the aspect ratio is not too high [35].

| Energy [eV] | Efficiency [%] |
|---|---|
| 30 | 7.6 |
| 100 | 7.6 |
| 450 | 8.2 |
| 900 | 8.1 |
| 1200 | 9.7 |
| 1500 | 16.0 |
| 2000 | 21.9 |
| 6000 | 11.2 |
| 8000 | 7.3 |

As for chromaticity, FZPs are chromatic optical devices and the placement of their zones is energy dependent [1]. Nevertheless, the energy dependence is small. In practice the design of a FZP with 100 nm outermost zone width and 100 µm diameter remains essentially unchanged from 100 eV to 8000 eV with difference in zone placement below 1.5 Å. Such small differences in zone placements can be neglected [36], the present FZPs can be employed to focus radiations from 100 eV to 8000 eV. Below 100 eV differences in zone placements are not negligible anymore and specific FZPs are to be prepared following the same preparation process.

Interestingly, while the intensity of light coming into even diffraction orders for standard rectangular FZPs is zero, it substantially increases if the FZP profile deviates from the standard shape or placement. As the resolution increases with the diffraction order (Equation (1)), this effect can be used to perform microscopy at higher resolutions [37-38]. Previous studies have shown for instance that by changing the line to space ratio of the zones from the standard value 1:1 to 1:3, the efficiency of the $2^{nd}$ order increases from zero up to a fourth of the $1^{st}$ order efficiency[38-39]. Nevertheless, though attractive, this effect is barely employed probably because resulting efficiencies are comparatively modest and because the required zone placement accuracy increases linearly with the diffraction order $m$ [36], which in turn imply that for an effective utilization of the $2^{nd}$ order imaging highly precise nanofabrication methods are required.

In the case of the present FZP, the trapezoidal profile leads to the activation of a $2^{nd}$ order of diffraction. The theoretical resolutions and diffraction efficiencies were estimated within the framework of the Takagi-Taupin dynamical diffraction theory [40] [Fig. 2]. The model takes possible diffraction phenomena within the volume of the zone into account as they may occur at the edges of the trapezoid. Nevertheless, results showed that for this structure dynamical effects are very weak. At 900 eV in the $1^{st}$ order focus of the FZP with a trapezoidal profile ($\alpha=6°$) the model predicts an overall efficiency of 7.8% which is very similar to the 8.1% predicted by the thin grating approximation (Table 1). In the $2^{nd}$ order the Takagi-Taupin model foresee an efficiency of 2.05% for $\alpha=6.0°$ and 2.24% for $\alpha=6.5°$ at 900 eV.

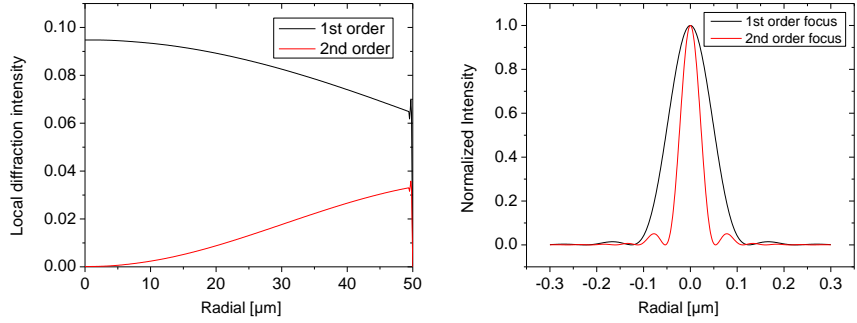

Fig. 2. Local diffraction efficiency as function of the FZP radius (left) and normalized intensity in the 1$^{st}$ and 2$^{nd}$ order focal plane, where the position of the first minimum is the resolution according to the Rayleigh criterion (right) for a trapezoidal FZP with α=6°. Calculation based on the Takagi-Taupin dynamical diffraction theory.

The utilization of the present FZP which is a positive FZP with a number of zones of 251 instead of negative FZPs which are usual should have no impact on the resolution as for a number of zones higher than 200 the resolution of both positive and negative FZPs approaches the resolution of a simple lens [41] and are given by Eq.(1).

## 4. Results

### a. Characterization of the FZP

The morphology of the FZP has been investigated by scanning electron microscopy (SEM) [Fig. 3(a) and (b)] and focused ion beam microscopy on a cross section [Fig. 3(c)].

Fig. 3(a) is an overview image of the FZP. A few randomly distributed gold particles in the spaces of the FZP are detected [Fig. 3(a)]. They are a result of the polycrystalline nature of the gold layer as grains with specific crystallographic orientations are more resistant to ion beam milling than others. As randomly distributed, the presence of the grains should slightly decrease the efficiency of the FZP but should not affect its resolution. Figure 2(b) confirms the overall period Λ of 200 nm for the outermost zones structures. Note that the gold lines in Fig. 3(b) appear larger than the spaces due to electron scattering on the trapezoidal form of the zones. Fig. 3(c) shows a cross section of the outermost zones. The trapezoidal form of the FZP is characterized by a top width of about 50 nm and a bottom width of about 150 nm and 50 nm empty spaces which corresponds to an inclination of the walls of α = 6°. A cross section in the center of the FZP (not shown) demonstrates a slightly lower inclination of the walls of the innermost zones with α = 4 to 5°. The roundness of the FZP was measured in a EBL tool equipped with an interferometer stage and dedicated metrology functionalities (RAITH150$^{TWO}$) [31]. The difference in FZP diameter in horizontal and vertical directions has been found to be below 25 nm which is low enough for astigmatism free imaging up to the 3$^{rd}$ order of diffraction (tolerable diameter difference $1.4\Delta r/m$ calculated similarly to [25, 36]).

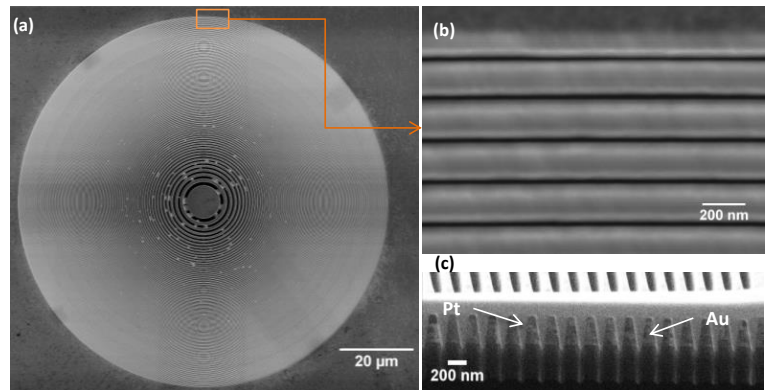

Fig. 3 Images of the FZP (a) SEM image recorded with an Everhart Thornley Detector (ETD): overview of the FZP manufactured according to the described procedure and further employed to perform imaging (b) SEM image recorded with a Through Lens Detector (TLD): closed view of the outermost zones, showing the 200 nm period ($\Lambda$) of the structure. (c) Ion Beam Image recorded at 45° on a cross section of a FZP manufactured under the same condition than the FZP in (a) and (b) and showing the trapezoidal from of the zones (prior to the cross-sectioning the zones are over coated with Pt to ensure their protection).

b. Imaging performance of the FZP

The focusing performances of the FZP were tested several times at several months intervals on two identical FZPs prepared according to the same procedures at the scanning transmission X-ray microscope (STXM) MAXYMUS located at the undulator UE46 PGM2 at BESSYII, HZB in Berlin [42]. The Beamline was operated with a 600l/mm grating, delivering high photon flux at a reasonable energy resolution. The employed fixfocuskonstante of the monochromator was $c = 1.8$. The FZP has been tested at various energies ranging from 450 eV to 1500 eV. At energies above 500 eV the undulator was operated in the $3^{rd}$ harmonic with a linear horizontal polarization while at lower energies below 500 eV the $1^{st}$ order harmonic and a circular polarization was used. This was necessary to avoid higher harmonic light contaminations at low energies. The amount of higher energy contamination was determined by measuring the absorption of Helium gas between 400 and 900 eV (absorption edge at about 867 eV). After correct alignment and configuration of the whole beamline contamination ratios with light of twice energy were measured to be of about 4%. In case of an improper alignment or configuration, as for instance an improper undulator-gap to monochromator alignment, contaminations up to 20% could be detected. As the FZP is mounted on a 500 nm Silicon Nitride membrane acting as a low energy filter, such contaminations may have dramatic impacts on the results. The full illumination of the FZP was realized by a divergent beam emerging from a 15 x 15 µm² (HxV) slit situated at the exit of the beamline 3 m upstream from the FZP. An order selecting aperture (OSA) of 35 µm diameter was placed between the FZP and the sample to eliminate higher diffraction orders and a part of the zero order diffraction radiation. Images of a commercially available test object, a Siemens Star (X-30-30-1, Xradia USA) with smallest feature size 30 nm were acquired at 1500, 1200, 900, 600 and 450 eV in the first, second and third order focus of the zone plate. Images of a second test object prepared by mean of a focused ion beam as a 500 nm cross section from a commercial BAM certified multilayer (L200) were recorded at 1200 eV with a beam slit size of 10 x 10 µm² (HxV) at the $1^{st}$, $2^{nd}$ and $3^{rd}$ order focus of the FZP to determine its ultimate resolution. The acquired images as well as the employed dwell times and the cutoff half-pitch resolutions determined from the power spectra of the respective images according to the procedure described in [43] are presented in Fig. 4, Fig. 5 and Table 2.

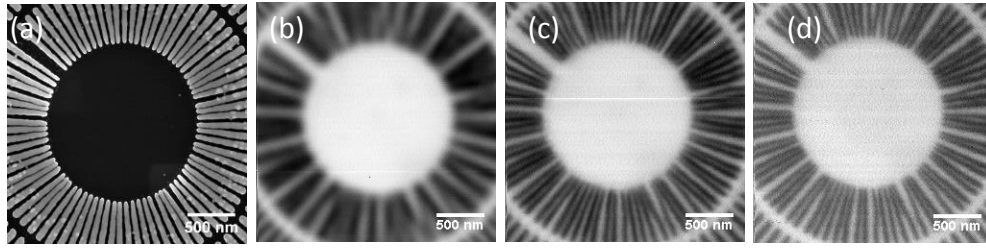

Fig. 4 (a) SEM image of a Siemens-star test object (b) (c) (d) Scanning X-ray microscopy of same test object at the 1st 2nd and 3rd order of diffraction acquired at 900 eV with a pixel size 10 nm and a dwell time of 10ms. The half pitch resolution was determined from the power spectrum of these images and are collected in Table 2

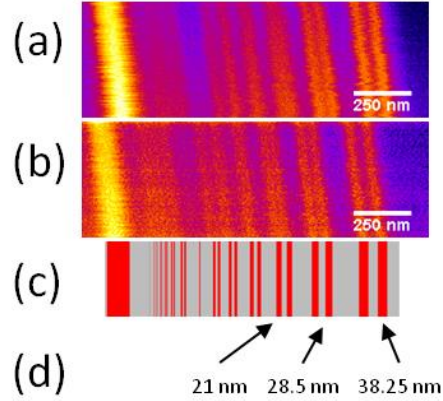

Fig. 5 Scanning X-ray microscopy image of a certified commercial test sample (BAM L200) recorded at 1200 eV (a) image acquired in the 2nd order focus of the FZP with a pixel size of 5 nm and a dwell time of 10 ms (b) image acquired in the 3rd order focus of the FZP with a pixel size of 5 nm and a dwell time of 15 ms (c) schematic representation of the certified test object (d) width of the features (half pitch)

Table 2  Spatial resolution performances obtained with the FZP in the 1st, 2nd and 3rd order focus. Where $\eta$ is the efficiency $\delta_{Ray}/2$ is the half-pitch (half-period) Rayleigh resolution and $\delta_{cutoff}/2$ is the half-pitch cutoff resolution determined from the power spectra of the images [Fig. 4(a) to (c)] and visually from the images [Fig. 5].

|  | Theoretical | Measured |
|---|---|---|
| Diffraction order | $\delta_{Ray}/2^{(a)}$ [nm] | $\delta_{cutoff}^{(b)}/2$ [nm] |
| 1st order | 61.0 | 44 [c] |
| 2nd order | 30.5 | 26 [c] |
| 3rd order | 20.3 | 21 [d] |

[a] Calculated according to Eq.(1) [b] the half pitch cutoff resolution ($\delta_{cutoff}/2$) is defined as the feature size below which no modulation can be detected, it is comparable to the half pitch Sparrow resolution limit [44] which is equal to $\delta_{Spa}/2 = 0.475\Delta r$ it is hence possible to detect features smaller than the Rayleigh resolutions (c) determined from the power spectra corresponding to the images in Fig. 4(d) visually determined from the images and from the known certified values in Fig. 5.

The efficiency of the FZP in the first diffraction order at various energies and in the 2nd diffraction order at 900 eV was determined experimentally as the ratio of light impinging on the FZP to the light focused by the FZP onto the detector through a 35 µm OSA, corrected for the 0th and higher order radiation according to the procedure described in [25]. The efficiencies at the first order focus are given in Fig. 6. The efficiency of the second diffraction order focus was measured to be 2.2% at 900 eV, which is in very good agreement with the efficiencies predicted by the Takagi-Taupin dynamical diffraction theory. This is in turn a sign of the high accuracy of the zone placement and of the fabrication. This is equally well illustrated by the high quality and contrast of the images obtained by X-ray scanning microscopy [Fig. 4 and Fig. 5].

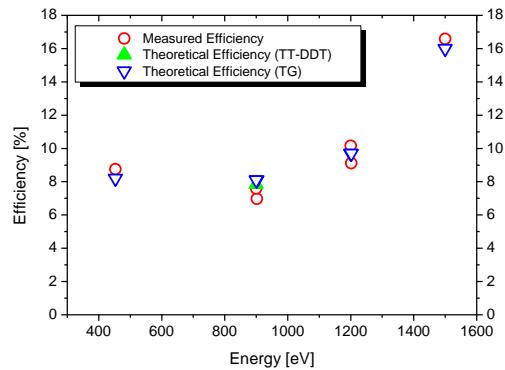

Fig. 6 Measured and theoretical efficiencies for the first order focus of the FZP. The theoretical efficiencies for a trapezoidal FZP were calculated for an angle α = 6° within the frame work of the thin grating theory (TG) and the Takagi-Taupin dynamical diffraction theory (TT-DDT). Note that in the real structure the angle α takes value between 4° and 6° which could explain the differences between theoretical and measured values.

## 5. Conclusion

Ion Beam Lithography (IBL) can be used to produce FZPs for X-ray radiation. The method enables FZPs in one single step. Compared with standard FZP manufacturing methods such as those based on EBL, it presents a number of advantages. The number of required manufacturing tools, equipment and skills is substantially decreased as well as the accumulation of errors typical of multistep processes. The achieved FZP thickness of 500 nm makes the method appropriate for the preparation of FZP for a wide range of energy ranging from the extreme UV to the limit of the hard X-ray range from 30 eV to 8000 eV with good efficiencies. The 1st order efficiency at 1500 eV has been measured to be above 16%. As the method inherently delivers FZP with trapezoidal profiles, this activates the 2nd order focus of the FZP and facilitates imaging resolutions twice as high as in the first order. Overall, the achieved resolutions allows the visualization of features of 61, 31 and 21 nm in size in the 1st, 2nd and 3rd order respectively and the corresponding measured diffraction efficiencies at the 1st and 2nd orders are very close to the theory. These performances were reproduced several times with an interval of several months using two different FZPs and showed a very good reproducibility. This shows the high reliability and the high quality and precision of IBL for the fabrication of FZPs in general and of the new *ionLINE®* tool in particular. Moreover, it is interesting to note that as ions are much heavier and less prompt to forward and backward scattering than electrons, the achievable feature size should essentially be determined by the beam size and its interaction with the milled material [26]. Consequently the potential resolution of IBL should even be higher than EBL. In fact, for various reasons such as shot noise effects, damage generations [45], limitations of usual IBL tools and inherent re-deposition phenomena [28, 33], the achieved resolutions to date remain lower in practice. Despite those apparent current drawbacks and issues, IBL shows a number of interesting characteristics or advantages over EBL and experienced interesting developments in the last years. It is a direct and mask-less nanofabrication technique and it allows the fast conception of complex structures [27, 46] moreover IBL can be applied to a large range of materials so that the method enables FZP made from other type of materials. With the increased interest in the last few years and the appearance of dedicated IBL devices [31, 45], for which the *ionLINE®* is an illustration, IBL may be considered as a serious alternative and solution towards higher resolutions lithography [47]. In particular, several studies and reviews have been published reporting about the ultimate resolution potential of IBL with for instance, the patterning of sub 10 nm features in HSQ resist with 30 nm periodicity [48], or the sub 5 nm nanopores in SiC membranes [46]. As for FZPs, studies towards a further simplification of the method and improved resolutions are underway. Considering the emergence of new X-ray sources and the related potential further expansion of X-ray microscopy both at large facilities and in the laboratories [5, 7-13], IBL represent a promising alternative for the simplified preparation of Fresnel Zone Plates.

**Acknowledgements**

The Authors thank Ulrike Eigenthaler and Dr. Michael Hirscher (Max-Planck-Institute for Intelligent Systems) for their support with the DualBeam, as well as Bernd Ludescher (Max-Planck-Institute for Intelligent Systems) for the gold layer preparation. In addition, we want to thank Dr. Hermann Stoll (Max-Planck-Institute for Intelligent Systems) and Tolek Tyliszczak (Advanced Light Source, Lawrence Berkeley National Laboratory) for the fruitful discussions.